%
%


\magnification=1200

\font\gross=cmbx10 scaled \magstep2
\font\mittel=cmbx10 scaled\magstep1
\font\pl=cmssq8 scaled \magstep1\def\RR{\rm I\!R}
\font\sc=cmcsc10
\font\sf=cmss10
\def\h#1{{\cal #1}}
\def\a{\alpha}
\def\b{\beta}
\def\g{\gamma}
\def\d{\delta}

\def\l{\lambda}
\def\m{\mu}
\def\n{\nu}

\def\om{\omega}
\def\na{\nabla}
\def\R{{\cal R}}

\def\sq{\Square}
\def\square#1{\mathop{\mkern0.5\thinmuskip\vbox{\hrule
    \hbox{\vrule\hskip#1\vrule height#1 width 0pt\vrule}\hrule}
    \mkern0.5\thinmuskip}}
\def\Square{\mathchoice{\square{6pt}}{\square{5pt}}
    {\square{4pt}}{\square{3pt}}}

\def\rightheadline{\it\author:\ \title\qquad\hfill\rm\folio}
\def\leftheadline{\it\author:\ \title\qquad\hfill\rm\folio}
\headline={\ifnum\pageno>1{\ifodd\pageno\rightheadline\else\leftheadline\fi}
\else\fi}

\def\author{Ivan G. Avramidi}
\def\title{Nonperturbative methods for calculating the heat
kernel}

%

\nopagenumbers

{\null
\vskip-1.5cm
\hskip5cm{ \hrulefill }
\vskip-.55cm
\hskip5cm{ \hrulefill }
\smallskip
\vskip1mm
\hskip5cm{{\pl \ University of Greifswald (January, 1996)}} 
\smallskip
\hskip5cm{ \hrulefill }
\vskip-.55cm
\hskip5cm{ \hrulefill } 
\bigskip 
\hskip5cm{\ hep-th/9602169}
\bigskip
\hskip5cm{\ To appear in: }
\smallskip
\hskip5cm{\ \sf Algebras, Groups and Geometries}

\vfill

\centerline{\gross Nonperturbative methods}
\medskip
\centerline{\gross for calculating the heat kernel}
\medskip
\vskip12pt
\centerline{\sc Ivan G. Avramidi
\footnote{*}{\rm Alexander von Humboldt Fellow. 
On leave of absence from Research Institute for Physics,
Rostov State University, Stachki 194, Rostov-on-Don 344104, Russia.}
\footnote{\dag}{AMS Subject Classification: 58G18, 58G11, 58G26, 81T20, 
81Q20}}
\bigskip
\centerline{\sl Department of Mathematics, University of Greifswald}
\centerline{\sl Jahnstr. 15a, 17487 Greifswald, Germany}
\centerline{\sl E-mail: avramidi@math-inf.uni-greifswald.d400.de}
\vfill

{\narrower 
\centerline{\sc Abstract}
\bigskip
We study the low-energy approximation for calculation of the heat kernel 
which is determined by the strong slowly varying background 
fields in strongly curved quasi-homogeneous manifolds. 
A new covariant algebraic approach, based on taking into account 
a finite number 
of low-order covariant derivatives of the background fields and 
neglecting all covariant derivatives of higher orders, is proposed.
It is shown that a set of covariant differential operators together 
with the background fields and their low-order derivatives generates 
a finite dimensional Lie algebra.
This
algebraic structure can be used to present the heat semigroup operator
in the form of an average over the corresponding Lie group. 
Closed covariant formulas for the
heat kernel diagonal are obtained.
These formulas serve, in particular, as the generating functions for the 
whole sequence 
of the Hadamard-\-Minakshisundaram-\-De~Witt-\-Seeley coefficients 
in all symmetric spaces.
\par}
\bigskip
\eject


\centerline{\mittel 1. Introduction}
\bigskip

The heat kernel is an important tool in quantum field theory and 
mathematical physics 
[1-13]. 
In particular, the one-loop contribution of quantized bosonic fields 
on a Riemannian manifold $(M,g)$ to the
effective action is given by the functional determinant of some elliptic 
differential operator $\Delta$ 
acting on the smooth sections $\varphi\in C^\infty(V)$  of a vector 
bundle $V$ over the manifold $M$ 
[1] 
$$
\Gamma_{(1)}={1\over 2}\log{\rm Det}\,\Delta.
\eqno(1.1)
$$
Using the standard spectral $\zeta$-function
[14], 
$$
\zeta(p)={\rm Tr}\, \Delta^{-p},
\eqno(1.2)
$$
where 'Tr' means the functional trace, one can present the effective 
action in terms of the derivative of the zeta-function at $p=0$:
$$
\Gamma_{(1)}=-{1\over 2}\zeta'(0).
\eqno(1.3)
$$
The most convenient way to evaluate the zeta-function in general 
case is to express it in terms of the corresponding heat kernel 
$\exp(-t \Delta)$
[15] 
$$
\zeta(p)={1\over \Gamma (p)}
\int\limits_0^\infty dt\ t^{p-1} 
\int\limits_M d\,{\rm vol}\;{\rm tr}\, U(t), 
\eqno(1.4)
$$
where 'tr' means the usual bundle (matrix) trace and $U(t)$ 
is the heat kernel diagonal, i.e. the restriction to the 
diagonal $M\times M$ of the heat kernel,  
$$
U(t)=\exp(-t \Delta)\delta(x,x')\big\vert_{x=x'},
\eqno(1.5)
$$
where $\delta(x,x')$ is the covariant distribution on 
$M\times M$.

By choosing the appropriate gauge and parametrization one 
can almost always reduce the problem to the Laplace type operators,
$$
\Delta=-\sq+Q+m^2,
\eqno(1.6)
$$
where $\sq=g^{\m\n}\na_\m\na_\n$ is the generalized Laplacian, 
$\nabla$ is
the covariant derivative on $C^\infty(V)$, $Q$ is 
an arbitrary endomorphism of the vector bundle $V$ and $m$ is 
a mass parameter. 

The effective action and the zeta function are determined 
by the spectrum of the operator $\Delta$ and are very 
complicated functionals of the background fields, i.e. 
the metric $g$, the bundle connection $\nabla$ and the 
endomorphism $Q$. 
Obviously, the effective action can be calculated exactly only 
for some very specific simple backgrounds. In quantum field 
theory, however, one needs the effective action for the 
{\it generic} background. Therefore, one has to develop 
consistent {\it approximate} methods for its calculation.
Moreover, these approximations should be manifestly 
{\it covariant}, i.e. they have to preserve the gauge 
invariance at each order. 

For the essentially {\it local} analysis that is carried out 
in this paper it is sufficient to characterize the background 
fields only by the local covariant objects, i.e. the curvatures, 
the Riemann curvature of $(M,g)$ and the curvature of the bundle 
connection $\nabla$, and their covariant derivatives. 
We denote the components of the Riemann curvature and the 
curvature of the bundle connection by $R_{\m\n\a\b}$ and 
${\cal R}_{\m\n}$ 
and call below all the quantities 
$\Re=\{R_{\m\n\a\b}, {\cal R}_{\m\n}, Q \}$ 
just the {\it background curvatures}. 

Further, following 
[16] 
we introduce the infinite set of all covariant derivatives 
of the curvatures, 
$$
{\cal J}=\{\Re_{(i)}; (i=1,2,\dots)\}, \qquad \Re_{(i)}=\{\underbrace{\nabla\cdots\nabla}_i \Re\}
\eqno(1.7)
$$
and call them the {\it background jets}. The whole set of jets, 
${\cal J}$ completely describes the background locally.

Since it is not possible to calculate the heat kernel exactly, 
one is forced to consider different asymptotic expansions.
A consistent way to construct the asymptotic expansions was 
developed in 
[16]. 
The idea is the following. One makes a deformation of the 
background fields with two deformation parameters, $\alpha$ 
and $\epsilon$,
$$
g\to g(\alpha,\epsilon), \qquad \nabla\to \nabla{(\alpha,\epsilon)},
\qquad Q\to Q(\alpha,\epsilon),
\eqno(1.8)
$$
in such a way that the jets transform uniformly,
$$
\Re_{(i)}\to \alpha\epsilon^i\Re_{(i)}.
\eqno(1.9)
$$

This deformation changes the operator $\Delta$ and, of course, 
the heat kernel $U(t)$,
$$
U(t)\to U(t;\alpha,\epsilon),
\eqno(1.10)
$$
and is manifestly covariant because of the transformation 
law (1.9). 

Thus it gives a natural framework to develop various asymptotic 
expansions with respect to the parameter $t$ and the deformation 
parameters $\a$ and $\epsilon$. The limit $t\to 0$ corresponds 
to small background jets, $t^{1+i/2}\Re_{(i)}\ll 1$, 
the limit $\alpha\to 0$ corresponds to the situation when the 
powers of curvatures are much smaller than the derivatives of 
them, so called short-wave, or {\it high-energy approximation}, 
$\na\na\Re\gg\Re\Re$, and the limit $\epsilon\to 0$ corresponds 
to the case when the derivatives of the curvatures are much 
smaller than the products of the curvatures of corresponding 
dimension, so called long-wave, or {\it low-energy approximation}, $\na\na\Re\ll\Re\Re$. For a more detailed discussion see 
[16-22]. 

As $t\to 0$, one has the well known asymptotic expansion
[1-9] 
$$
U(t)
\sim(4\pi t)^{-d/2}\exp(-tm^2)
\sum\limits_{k=0}^\infty{(-t)^k\over k!} b_k,
\eqno(1.11)
$$
where $d$ is the dimension of the manifold.
The coefficients $b_k$ are the famous 
Hadamard-\-Minak\-shi\-sun\-da\-ram-\-De~Witt-\-Seeley 
(HMDS) coefficients
[1,4-9,23-28]. 
They are purely local universal invariants built from the 
background curvatures and their covariant derivatives
that do not depend on the global structure of the manifold 
and the boundary conditions. 
They play a very important role both in physics and mathematics
[6,27,28]. 
The HMDS-coefficients are known now in general case up to $b_4$
[4,5,23-26]. 
The $b_4$ coefficient in general case was calculated for the 
first time in our PhD thesis 
[4] 
and has been published then in
[24,25,5]. 

However, the asymptotic expansion (1.11) is of very limited 
applicability.
It is absolutely inadequate for large $t$  ($t\Re \gg 1$) in 
strongly curved manifolds and strong background fields. 
Therefore, this approximation cannot describe essentially 
nonperturbative {\it nonlocal} and {\it nonanalytical} effects. 
The investigation of such effects requires consideration of 
other approximation schemes.

In the high-energy limit, $\a\to 0$, there is an expansion
[16] 
$$
U(t;\alpha,\epsilon)
\sim(4\pi t)^{-d/2}\exp(-tm^2)
\sum\limits_{n=0}^\infty (\a t)^n h_n(t;\epsilon),
\eqno(1.12)
$$
where $h_n(t;\epsilon)$ are some {\it nonlocal} functionals.
They are {\it resummed perturbative} objects.
This approximation was studied in details in our PhD thesis 
[4] 
and the papers
[5,29,30], 
where the explicit form of the functionals $h_1$ and $h_2$ 
was obtained and analyzed. 
The third coefficient $h_3$ has been investigated in 
[31]. 

The long-wave (or low-energy) approximation is determined by 
strong slowly varying  background fields.  
It corresponds to the asymptotic expansion of the deformed 
heat kernel as $\epsilon\to 0$
[16] 
$$
U(t;\alpha,\epsilon)
\sim(4\pi t)^{-d/2}\exp(-tm^2)
\sum\limits_{l=0}^\infty (\epsilon^2 t)^l u_l(t;\a).
\eqno(1.13)
$$
The coefficients $u_l$ are essentially {\it non-perturbative} 
functionals.
They cannot be obtained in any perturbation theory and are much 
complicated than the HMDS-coefficients $b_k$ and the high-energy 
functionals $h_n$.

We consider in this paper mostly
the zeroth order of this approximation, i.e. the coefficient $u_0$, 
which corresponds simply to covariantly constant background curvatures
$$
\na_\m R_{\a\b\g\d} = 0,\qquad \na_\m{\cal R}_{\a\b}=0,
\qquad \na_\m Q = 0.
\eqno(1.14)
$$
The coefficient $u_0$ depends, of course, on the global structure 
of the manifold. However, the asymptotic expansion of $u_0$ as 
$t\to 0$ is purely local and determines all the terms without 
covariant derivatives in all HMDS-coefficients $b_k$. 
 Therefore, it can be viewed on as the 
{\it generating function} for all HMDS-coefficients in covariantly 
constant background.

The conditions (1.14) determine the geometry of {\it locally} 
symmetric spaces
[32,33]. 
The {\it globally} symmetric manifold satisfies additionally some 
topological restrictions and the condition (1.14) is valid everywhere 
in the manifold.
However, in typical {\it physical } problems, the situation is rather
different. One has usually a complete noncompact asymptotically flat 
space-time
manifold without boundary that is homeomorphic to $\RR^d$. 
In the low-energy approximation a  finite
(not small, in general) region of the manifold exists that is locally 
strongly curved
and quasi-homogeneous, i.e., the local invariants of the curvature 
in this region
vary very slowly.  Then the geometry of this region is {\it locally} 
very similar
to that of a symmetric space.  However, globally the manifold can 
be completely different from the symmetric space and one should 
keep in mind that there are
{\it always} regions in the manifold where the condition (1.14) 
is not fulfilled.
See the discussion in 
[17-22]. 

Thus the problem is to calculate  the low-energy heat kernel 
diagonal (1.5) 
for covariantly constant background (1.14). In other words one 
has to construct a local
covariant function of the invariants of the curvatures that
would describe adequately the low-energy limit of the heat
kernel diagonal and that would, when expanded in curvatures, 
reproduce {\it all terms without covariant derivatives} in the 
asymptotic expansion of the heat kernel. If one finds such an 
expression, then one can simply determine the $\zeta$-function, 
(1.4), and, therefore, the effective action, (1.3).


\bigskip
\bigskip
\centerline{\mittel 2. Algebraic approach}
\bigskip

There exist a very elegant indirect way to construct the heat 
kernel without solving the heat equation but using only the 
commutation relations of some covariant first order differential 
operators
[16-22]. 
The main idea is in a generalization of the usual Fourier 
transform to the case of operators and consists in the following.

Let us consider for a moment a trivial case, where the 
curvatures vanish but not the potential term:
$$
R_{\a\b\g\d} = 0,\qquad {\cal R}_{\a\b}=0,\qquad \nabla Q=0.
\eqno(2.1)
$$
In this case the operators of covariant derivatives obviously 
commute and form together with the potential term an Abelian 
Lie algebra
$$
[\na_\m,\na_\n]=0,\qquad [\na_\m, Q]=0. 
\eqno(2.2)
$$
It is easy to show that the {\it heat semigroup operator} can 
be presented in the form
$$
\eqalignno{
\exp(-t \Delta)=&(4\pi t)^{-d/2}\exp[-t(m^2+Q)]&\cr
&\times\int\limits_{\RR^d} dk g^{1/2}
\exp\left(-{1\over 4t}<k,g k>
+k\cdot\na\right),&
(2.3)\cr}
$$
where $<k,g k>=k^\m g_{\m\n}k^\n$, $k\cdot\nabla=k^\m \na_\m$ 
and $g^{1/2}=\sqrt{\det g_{\m\n}}$.
Here, of course, it is assumed that the covariant derivatives 
commute also with the metric
$$
[\na_\m,g_{\a\b}]=0. 
\eqno(2.4)
$$
Acting with this operator on the $\d$-function and using the 
obvious relation
$$
\exp(k\cdot\na)\d(x,x')\big\vert_{x=x'}=g^{-1/2}\d(k),
\eqno(2.5)
$$
one integrates easily over $k$ and obtains the heat kernel 
diagonal
$$
U(t)=(4\pi t)^{-d/2}\exp[-t(m^2+Q)].
\eqno(2.6)
$$

In fact, the commutators of the covariant differential operators 
$\na$ do not vanish but are proportional to the curvatures $\Re$. 
The commutators of covariant derivatives $\na$ with the curvatures 
$\Re$ give the first 
derivatives of the curvatures, i.e. the jets $\Re_{(1)}$, the 
commutators of 
 covariant derivatives with $\Re_{(1)}$ give the second jets 
 $\Re_{(2)}$, etc.
Thus the operators $\na$ together with the 
whole set of the jets $\h J$ form an {\it infinite} dimensional 
Lie algebra 
$\h G=\{\na, \Re_{(i)}; (i=1,2,\dots)\}$
[16]. 
To evaluate the low-energy heat kernel one 
can take into account a {\it finite} number of low-order jets, 
i.e. the 
low-order covariant derivatives of the background fields,
$\{\Re_{(i)}; (i\le N)\}$,  and neglect all 
the higher order jets, i.e. the covariant derivatives of 
higher orders,
i.e. put $\Re_{(i)}=0$\ for $i> N$.
Then one can show that there exist a set of covariant differential 
operators 
that together with the background fields and their low-order 
derivatives 
generate a {\it finite} dimensional Lie algebra 
$\h G'=\{\nabla, \Re_{(i)};
(i=1,2,\dots,N)\}$
[16]. 

Thus one can try to generalize 
the above idea in such a way that (2.3) would be the zeroth 
approximation 
in the commutators of the covariant derivatives, i.e. in the 
curvatures. 
Roughly speaking, we are going to find a representation of the 
heat semigroup 
{\it operator} 
in the form
$$
\exp(-t \Delta)=
\int\limits_{\RR^D} dk\,
\Phi(t,k)\exp\left(-{1\over 4t}<k,\Psi(t) k>
+k\cdot T\right)
\eqno(2.7)
$$
where $<k,\Psi(t) k>=k^A\Psi_{AB}(t)k^B$, 
$k\cdot T=k^A T_A$, ($A=1,2,\dots,D$), 
 $T_A=X^\m_A\na_\m+Y_A$ are some first order differential operators 
 and the functions $\Psi(t)$ and $\Phi(t,k)$ are expressed in 
 terms  of commutators of these 
operators--- i.e., in terms of the curvatures.

In general, the operators $T_A$ do not form a closed finite 
dimensional algebra because at each stage taking more 
commutators there appear more and more derivatives of the 
 curvatures. It is the {\it low-energy reduction} 
${\cal G}\to {\cal G}'$, i.e. the restriction to the 
low-order jets, that actually closes the algebra ${\cal G}$ 
of the operators $T_A$ and the background jets, i.e. makes 
it finite dimensional.

Using this representation one could, as above, act with 
$\exp(k\cdot T)$ on the $\d$-function on $M$ to get the 
heat kernel. The main point of this idea is that it is 
much easier to calculate the action of the exponential 
of the {\it first} order operator $k\cdot T$ on the 
$\d$-function than that of the exponential of the second 
order operator $\sq$.


\bigskip
\bigskip
\centerline{\mittel 3. Heat kernel in flat space}
\vglue0pt
\smallskip
\vglue0pt
\centerline{\bf 3.1 Covariantly constant potential term}
\vglue0pt
\bigskip
\vglue0pt
Let us consider now the more complicated case of nontrivial 
covariantly constant curvature of background connection in 
flat space:
$$
R_{\a\b\g\d} = 0,\qquad \na_\m{\cal R}_{\a\b}=0,
\qquad \na_\m Q=0.
\eqno(3.1)
$$

Using the condition of covariant constancy of the curvatures (1.14)
one can show that in this case the covariant derivatives form 
a {\it nilpotent} Lie algebra
$$
\eqalignno{
&[\na_\m,\na_\n]=\h R_{\m\n}, &(3.2)\cr
&[\na_\m,\h R_{\a\b}]=[\na_\m,Q]=0, &\cr
&[\h R_{\m\n},\h R_{\a\b}]=[\h R_{\m\n},Q]=0.\cr}
$$

For this algebra one can prove a theorem expressing the heat 
semigroup operator in terms of an average over the 
corresponding Lie group
[17,18] 
$$
\eqalignno{
\exp(-t \Delta)&=(4\pi t)^{-d/2}\exp[-t(m^2+Q)]
\det\left({t\h R\over \sinh(t\h R)}\right)^{1/2}&\cr
&\times\int\limits_{\RR^d}dk g^{1/2}
\exp\left(-{1\over 4t}<k, g t\h R \coth(t\h R) k>
+k\cdot\na\right),
&(3.3)\cr}
$$
where $k\cdot\na=k^\m\na_\m$, $\h R$ means the matrix with 
coordinate indices $\h R=\{\h R^\m_{\ \n}\}$, 
$\h R^\m_{\ \n}=g^{\m\l}\h R_{\l\n}$,  and the 
determinant is taken with respect to these indices, 
other (bundle) indices being intact.

It is not difficult to show that also in this case we have
$$
\exp(k\cdot\na)\d(x,x')\big\vert_{x=x'}
=g^{-1/2}\d(k).
\eqno(3.4)
$$
Subsequently, the integral over $k^\m$ becomes trivial and 
we obtain immediately the heat kernel diagonal
$$
U(t)=(4\pi t)^{-d/2}\exp[-t(m^2+Q)]
\det\left({t\h R \over \sinh(t\h R)}\right)^{1/2}. 
\eqno(3.5)
$$
Expanding it in a power series in $t$ one can find {\it all} 
covariantly constant terms in {\it all} HMDS-coefficients $b_k$.

As we have seen the contribution of the bundle curvature 
$\h R_{\m\n}$ is not as trivial as that of the potential term. 
However, the algebraic approach does work in this case too. 
It is a good example how one can get the heat kernel without 
solving any differential equations but using only the algebraic 
properties of the covariant derivatives.

\bigskip
\bigskip
\centerline{\bf 3.2 Contribution of two first derivatives of 
the potential term}
\bigskip

In fact, in flat space it is possible to do a bit more, i.e. 
to calculate the contribution of the first and the second 
derivatives of the potential term $Q$
[22]. 
That is we consider the case when the derivatives of the 
potential term vanish only starting from the {\it third} 
derivative, i.e.
$$
R_{\a\b\g\d} = 0,\qquad \na_\m{\cal R}_{\a\b}=0,\qquad 
\na_\m\na_\n\na_\l Q=0.
\eqno(3.6)
$$
Besides we assume the background to be {\it Abelian}, 
i.e. all the nonvanishing background quantities, 
$\h R_{\a\b}$, $Q$, $Q_{;\m}\equiv\na_\m Q$ and 
$Q_{;\n\m}\equiv \na_\n\na_\m Q$, commute with each other. 
Thus we have again a nilpotent Lie algebra
$$
\eqalignno{
&[\na_\mu, \na_\nu]={\cal R}_{\mu\nu}, &\cr
&[\na_\m,Q]=Q_{;\m}, &(3.7)\cr
&[\na_\m,Q_{;\n}]=Q_{;\n\m}, &\cr}
$$
all other commutators being zero.

By parametrizing the potential term according to
$$
Q=\Omega-\a^{ik}N_iN_k,
\eqno(3.8)
$$
where $(i=1,\dots,q; q\le d),$ $\a^{ik}$ is some constant 
symmetric nondegenerate $q\times q$ matrix, $\Omega$ is a 
covariantly constant matrix and $L_i$ are some matrices with 
vanishing second covariant derivative:
$$
\na_\m \Omega=0, \qquad \na_\m\na_\n N_i=0,
\eqno(3.9)
$$
and introducing the operators $X_A=(\na_\m, N_i)$, 
$(A=1,\dots, d+q)$, 
one can rewrite the commutation relations (3.7) in 
a more compact form
[22] 
$$
\eqalignno{
&[X_A, X_B]=\h F_{AB}, &(3.10)\cr
&[X_A, \h F_{CD}]=[X_A, \Omega]=0,&\cr
&[\h F_{AB}, \h F_{CD}]=[\h F_{AB}, \Omega]=0,   &\cr}
$$
where $\h F_{AB}$ is a matrix
$$
(\h F_{AB})=\left(\matrix{
\R_{\m\n} & N_{i;\m}\cr
-N_{k;\n} & 0       \cr}\right),
\eqno(3.11)
$$
with $N_{i;\m}\equiv \na_\m N_i$.

The operator (1.6) can now be written in the form
$$
\Delta=-\l^{AB}X_A X_B+\Omega+m^2,
\eqno(3.12)
$$
where
$$
(\l^{AB})=\left(\matrix{g^{\m\n} & 0 \cr
			0 & \a^{ik} \cr}\right).
\eqno(3.13)
$$

The algebra (3.10) is a nilpotent Lie algebra of the type (3.2).
Thus one can apply the theorem (3.3) in this case too to get
[22] 
$$
\eqalignno{
\exp(-t &\Delta)=(4\pi t)^{-D'/2}\exp[-t(\Omega+m^2)]
\det\left({\sinh(t\h F)\over t\h F}\right)^{-1/2}
&\cr&\times\int\limits_{\RR^{d+q}} d k \l^{1/2}
\exp\left(-{1\over 4t}<k, \l t\h F \coth(t\h F)k>
+k\cdot X\right),
&(3.14)\cr}
$$
where $\l=\det\l_{AB}$, 
$<k, \l t\h F \coth(t\h F))k>
=k^A\l_{AB}(t\h F \coth(t\h F))^A_{\ C}k^C$,
$k\cdot X=k^A X_A$.

Thus we have expressed the heat semigroup operator in terms 
of the operator 
$\exp(k\cdot X)$. The integration over $k$ in (4.16) is 
Gaussian except 
for the noncommutative part.
Splitting the integration variables $(k^A)=(q^\m, \om^i)$ and 
using the 
Campbell-Hausdorf formula we obtain 
[22] 
$$
\exp(k\cdot X)\d(x,x')\Big\vert_{x=x'}
=g^{-1/2}\exp(\om\cdot N)\d(q),
\eqno(3.15)
$$
where $\om\cdot N=\om^i N_i$. 

Further, after taking off the trivial 
integration over $q$
and a Gaussian integral over $\om$,  
we obtain the heat kernel diagonal in a very simple form 
[22] 
$$
U(t)=(4\pi t)^{-d/2}\Phi(t) 
\exp\left[-t(m^2+Q)
+{1\over  4} t^3 <\na Q, \Psi(t)g^{-1}\na Q>\right],
\eqno(3.16)
$$
where $
<\na Q, \Psi(t)g^{-1}\na Q>
=\na_\m Q\Psi^\m_{\ \n}(t)g^{\n\l}\na_\l Q,
$
$$
\Phi(t)=\det\left({\sinh(t\h F)\over t\h F}\right)^{-1/2}
\det(1+t^2C(t)P)^{-1/2},
\eqno(3.17)
$$
$$
\Psi(t)=\{\Psi^\m_{\ \n}(t)\}=(1+t^2C(t)P)^{-1}C(t),
\eqno(3.18)
$$
$P$ is the matrix determined by second derivatives of the 
potential term,
$$
P=\left\{P^\m_{\ \n}\right\}, \qquad 
P^\m_{\ \n}={1\over 2} g^{\m\l}\na_\n \na_\l Q,
\eqno(3.19)
$$
 and
the matrix $C(t)=\{C^\m_{\ \n}(t)\}$ is defined by 
$$
C(t)=\oint\limits_C{dz\over 2\pi i}t 
\coth({tz^{-1}})(1-z\h R-z^2 P)^{-1}.
\eqno(3.20)
$$

The formula (3.16) exhibits the general structure of the heat 
kernel diagonal. 
Namely, one sees immediately how the potential term and its 
first derivatives 
enter the result. The complete nontrivial information is 
contained only in a 
scalar, $\Phi(t)$, and a tensor, $\Psi_{\m\n}(t)$, functions 
which are 
constructed purely from the curvature $\R_{\m\n}$ and the 
{\it second} derivatives of the potential term, $\na_\m\na_\n Q$. 
So we 
conclude that the coefficients $b_k$ of the heat kernel 
asymptotic expansion
(1.11) are 
constructed from three different types of scalar (connected) blocks, 
$Q$, $\Phi_{(n)}(\R, \na\na Q)$ and 
$\na_\m Q\Psi^{\m\n}_{(n)}(\R, \na\na Q)\na_\n Q$.


\bigskip
\bigskip
\centerline{\mittel 4. Heat kernel in symmetric spaces}
\vglue0pt\bigskip\vglue0pt
Let us now generalize the algebraic approach
to the case of the {\it curved} manifolds with covariantly constant 
Riemann curvature and the trivial bundle connection
[19,20]: 
$$
\na_\m R_{\a\b\g\d}=0, \qquad \R_{\a\b}=0, \qquad \na_\m Q=0.
\eqno(4.1)
$$

First of all, we give some definitions 
[32,33]. 
The condition (4.1) defines, as we already said above, 
the geometry of locally symmetric spaces. 
A Riemannian locally symmetric space which is simply connected and 
complete is
globally symmetric space (or, simply,  symmetric space).
A symmetric space is said to be of {\it compact, noncompact 
{\rm or} 
Euclidean type} if {\it all} sectional curvatures 
$K(u,v)=R_{abcd}u^av^bu^cv^d$ 
are positive, negative or zero. A direct product of symmetric 
spaces of 
compact and noncompact types is called {\it semisimple} 
symmetric space. 
A generic complete simply connected 
Riemannian symmetric space is a direct product of a flat space and a 
semisimple symmetric space. 

It should be noted that our analysis in this paper is purely 
{\it local}.  We are looking for a 
{\it universal} local function of the curvature invariants, 
$u_0$ (introduced in Sect.1)  that 
describes adequately  the low-energy limit of the heat kernel 
diagonal 
$U(t)$. 
Our minimal requirement is that this function should reproduce 
{\it all} the 
terms without covariant derivatives of the curvature in the local 
asymptotic expansion of the heat kernel (1.11), i.e. 
it should give {\it all} the HMDS-coefficients $b_k$  for {\it any} 
symmetric space.

It is well known that the HMDS-coefficients have a {\it universal} 
structure, i.e.
they are polynomials in the background jets (just in curvatures 
in case of symmetric spaces) 
with the numerical coefficients that do not depend
on the global properties of the manifold, on the dimension, on 
the signature of the metric etc.
[6]. 

It is obvious that any flat subspaces do not contribute to 
the HMDS-coefficients $b_k$. 
Therefore, to find this universal structure it is sufficient 
to consider only semisimple 
symmetric spaces. Moreover, since HMDS-coefficients are 
analytic in the curvatures, one can restrict oneself only 
to symmetric spaces of {\it compact} type.
Using the factorization property of the heat kernel and 
the duality 
between compact and noncompact symmetric spaces one can 
obtain then the 
 results for the general case by analytical continuation.
That is why in this paper we consider only the case of 
{\it compact} symmetric 
spaces  when the sectional curvatures and the metric
are {\it positive} definite. 

Let $e_a^\m$ be a {\it covariantly constant (parallel)} 
frame along the geodesic.
The frame components of the curvature tensor of a symmetric 
space are, obviously, constant and can be presented in the form 
[19,20] 
$$
R_{abcd} = \b_{ik}E^i_{\ ab}E^k_{\ cd}, 
\eqno(4.2)
$$
where $E^i_{ab}$, $(i=1,\dots, p; p \le d(d-1)/2)$, is some 
set of antisymmetric matrices and $\b_{ik}$ is some symmetric 
nondegenerate $p\times p$ matrix.
The traceless matrices $D_i=\{D^a_{\ ib}\}$ defined by
$$
D^a_{\ ib}=-\b_{ik}E^k_{\ cb}g^{ca}= - D^a_{\ bi} 
\eqno(4.3)
$$
are known to be the generators of the {\it holonomy algebra} 
${\cal H}$
$$
[D_i, D_k] = F^j_{\ ik} D_j, 
\eqno(4.4)
$$
where $F^j_{\ ik}$ are the structure constants. 

In symmetric spaces a much richer algebraic structure exists
[19,20]. 
Indeed, 
let us define the quantities $C^A_{\ BC}=-C^A_{\ CB}$, 
$(A=1,\dots, D; D=d+p)$:
$$
C^i_{\ ab}=E^i_{\ ab}, \quad C^a_{\ ib}
=D^a_{\ ib}, \quad C^i_{\ kl}=F^i_{\ kl}, 
\eqno(4.5)
$$
$$
C^a_{\ bc}=C^i_{\ ka}=C^a_{\ ik}=0,
$$
and the matrices $C_A=\{C^B_{\ AC}\}=(C_a,C_i)$:
$$
C_a = \left( \matrix{ 0          & D^b_{\ ai}   \cr
		      E^j_{\ ac} & 0            \cr}\right), 
\qquad
C_i = \left( \matrix{ D^b_{\ ia} & 0            \cr
		      0          & F^j_{\ ik}   \cr}\right).
\eqno(4.6)
$$
One can show that they satisfy the Jacobi identities
$$
[C_A, C_B]=C^C_{\ AB}C_C
\eqno(4.7)
$$
and, hence, define a Lie algebra ${\cal G}$ of dimension $D$ with 
the structure constants $C^A_{\ BC}$, the matrices
$C_A$ being the generators of adjoint representation.

In symmetric spaces one can find explicitly the generators 
of the infinitesimal isometries, i.e. the Killing
vector fields $\xi_A$,
and show that they form a Lie algebra of isometries that is 
(in case of semisimple symmetric space) isomorphic to the
Lie algebra ${\cal G}$ (4.7),
[20], 
viz.
$$
[\xi_A,\xi_B]=C^C_{\ AB}\xi_C.    
\eqno(4.8)
$$
Moreover, introducing a symmetric nondegenerate $D\times D$ 
matrix
$$
\g_{AB} = \left(\matrix{ g_{ab} & 0             \cr
	       0                & \b_{ik}        \cr}\right),
\eqno(4.9)
$$
that plays the role of the metric on the algebra ${\cal G}$, 
one can
express the operator (1.6) in semisimple symmetric spaces in
terms of the generators of isometries
[20] 
$$
\Delta=-\g^{AB}\xi_A\xi_B+Q+m^2,
\eqno(4.10)
$$
where $\g^{AB}=(\g_{AB})^{-1}$.

Using this representation one can prove a theorem that 
presents the heat 
semigroup operator in terms of some average over the 
group of 
isometries $G$
[19,20]: 
$$
\eqalignno{
\exp(-t &\Delta)=
(4\pi t)^{-D/2}\exp\left[-t(Q+m^2-{1\over 6} R_G)\right]&\cr
&\times\int\limits_{\RR^D} d k\g^{1/2}
\det\left({\sinh(k\cdot C/2)\over k\cdot C/2}\right)^{1/2}
\exp\left(-{1\over 4t}<k,\g k>
+k\cdot \xi\right)&\cr
& &(4.11)\cr}
$$
where $\g=\det\g_{AB}$, $k\cdot C=k^A C_A$, $k\cdot \xi=k^A\xi_A$, 
and $R_G$ is the scalar curvature of the group of isometries $G$
$$
R_G= -{1\over 4}\g^{AB} C^C_{\ AD}C^D_{\ BC}. 
\eqno(4.12)
$$

Acting with this operator on the delta-function $\d(x,x')$ one can,
in principle, evaluate the off-diagonal heat kernel 
$\exp(-t\Delta)\d(x,x')$, i.e. for non-coinciding points $x\ne x'$ 
[20]. 
Since in this paper we are going to calculate only the heat 
kernel diagonal (1.5),
it is sufficient to compute only the coincidence limit $x=x'$.  
Splitting the integration variables $k^A=(q^a,\om^i)$ and 
solving the equations of characteristics one can obtain the 
action of the isometries on the $\d$-function 
(Lemma 2 in [20]) 
$$
\exp\left(k\cdot\xi\right)\d(x,x')\Big\vert_{x=x'}
=\det\left({\sinh(\om\cdot D/2)\over 
\om\cdot D/2}\right)^{-1}\eta^{-1/2}\d(q).  
\eqno(4.13)
$$
where $\om\cdot D=\om^i D_i$ and $\eta=\det\,g_{ab}$.
Using this result one can easily integrate over $q$ in 
(4.11) to get the 
heat kernel diagonal. After changing the integration 
variables $\om \to \sqrt t \om$ it takes the form
$$
\eqalignno{
U(t)=&(4\pi t)^{-d/2}\exp\left[-t\left(m^2-{1\over 8}R
-{1\over 6} R_H\right)\right]& \cr
&\times
(4\pi)^{-p/2}\int\limits_{\RR^p}d \om\, \b^{1/2}
\exp\left(-{1\over 4} <\om,\beta \om>\right)&\cr
&\times\det\left({\sinh(\sqrt t \om\cdot F/2)
\over \sqrt t \om\cdot F/2}\right)^{1/2}
\det\left({\sinh(\sqrt t \om\cdot D/2)
\over \sqrt t \om\cdot D/2}\right)^{-1/2},
&(4.14)\cr}
$$
where $\om\cdot F=\om^i F_i$, $F_i=\{F^j_{\ ik}\}$ are 
the generators of the holonomy algebra ${\cal H}$, (4.4), 
in adjoint representation and
$$
R_H=-{1\over 4}\b^{ik}F^m_{\ il}F^l_{\ km}
\eqno(4.15)
$$
is the scalar curvature of the holonomy group.

The remaining integration over $\om$ in (4.14) can be 
done in a rather {\it formal} way.
Namely, one can prove that for any analytic function 
$f(\om)$  that falls off at the infinity there holds
[20] 
$$
\eqalignno{
(4\pi)^{-p/2}\int\limits_{\RR^p} d \om\, \b^{1/2}
\exp&\left(-{1\over 4}<\om,\b\om>\right)f(\om)&\cr
&=f\left(i{\partial\over \partial q}\right)
\exp\left(-<q,\b^{-1}q>\right)\Big\vert_{q=0},
&(4.16)\cr}
$$
where $<q,\b^{-1}q>=q_j\b^{jk}q_k$.

Introducing an abstract dynamical system with a 
normalized `vacuum state' $|0>$,
$$
<0|0>=1,
\eqno(4.17)
$$
and the `coordinate' and `momentum' operators 
$\hat q_i$ and $\hat p^k$ satisfying
the commutation relations
$$
[\hat p^j, \hat q_k]=i\d^j_{\ k},
\eqno(4.18)
$$
$$
[\hat p^i,\hat p^k]=[\hat q_i,\hat q_k]=0,
$$
and the rules
$$
\hat p^i|0>=0, \qquad <0|\hat q_k=0,
\eqno(4.19)
$$
one can present the equation (4.16) in the form
$$
\eqalignno{
(4\pi)^{-p/2}\int\limits_{\RR^p} d \om\, \b^{1/2}
\exp&\left(-{1\over 4}<\om,\b\om>\right)f(\om)&\cr
&=<0|f\left(\hat p\right)
\exp\left(-<\hat q,\b^{-1}\hat q>\right)|0>.
&(4.20)\cr}
$$
Using this equation we have finally from (4.14) 
the heat kernel diagonal in an formal algebraic form 
without any integration
$$
\eqalignno{
U(t)=&(4\pi t)^{-d/2}
\exp\left[-t\left(m^2+Q-{1\over 8}R
-{1\over 6}R_H\right)\right]&\cr
&\times\Big<0\Big|
\det\left({\sinh(\sqrt t \hat p\cdot F/2)\over 
\sqrt t \hat p\cdot F/2}\right)^{1/2}
\det\left({\sinh(\sqrt t \hat p\cdot D/2)\over 
\sqrt t \hat p\cdot D/2}\right)^{-1/2}
&\cr
&\times\exp\left(-<\hat q,\beta^{-1}\hat q>\right)\Big|0\Big>. 
&(4.21)\cr}
$$
where $\hat p\cdot F=\hat p^k F_k$ and 
$\hat p\cdot D=\hat p^k D_k$.
This formal solution should be understood as a power 
series in the operators $\hat p^k$ 
and $\hat q_k$ and determines a well defined asymptotic 
expansion in $t\to 0$.

Let us stress that the formulae (4.14) and (4.21) 
are exact (up to topological contributions) and 
{\it manifestly covariant} because they are expressed 
in terms of the invariants of the holonomy group $H$, 
i.e. the invariants of the Riemann curvature tensor. 
They can be used now to generate {\it all} HMDS-coefficients 
$b_k$ for {\it any} symmetric space, i.e. for any manifold 
with covariantly constant curvature, simply by expanding 
it in an asymptotic power series as $t\to 0$. 
Thereby one finds {\it all} covariantly constant terms in 
{\it all} HMDS-coefficients in a manifestly covariant way. 
This gives a very nontrivial example how the heat kernel 
can be constructed using only the Lie algebra of isometries 
of the symmetric space.


\bigskip
\bigskip
\centerline{\mittel 9. Conclusion}
\bigskip

In present paper we have presented our recent results  
in studying the heat kernel obtained in the papers 
[16-22]. 
We discussed some ideas connected with the problem of 
developing consistent covariant
approximation schemes for calculating the heat kernel.
Especial attention is payed to the low-energy limit of 
quantum field theory.
It is shown that  in the local analysis there exists 
an algebraic structure (the Lie algebra of background 
jets) that turns out to be extremely useful 
for the study of the low-energy approximation. Based 
on the background jets algebra 
we have proposed a new promising approach for  calculating 
the low-energy heat kernel.

Within this framework we have obtained closed formulas 
for the heat kernel diagonal in the zeroth order, i.e. 
in case of covariantly constant background curvatures. 
Besides, we were able to take into account the first 
and second derivatives of the potential term in flat 
space (Sect. 3.1). 

The obtained formulas are exact, covariant and general, 
i.e. they are applicable for {\it any} covariantly 
constant background fields. This enables to treat the 
results of this paper as the {\it generating functions} 
for the whole set of the 
Hadamard-\-Minakshisundaram-\-De~Witt-\-Seeley-\-coefficients. 
In other words, we have calculated {\it all} covariantly 
constant terms in {\it all} HMDS-\-coefficients. 
This is the opposite case to the leading derivatives 
terms which were calculated in 
[4,29,30,5] 
and 
[34]. 

Needless to say that the investigation of the low-energy 
effective action is of great importance in quantum gravity 
and gauge theories because it describes the dynamics of 
the vacuum state of the theory. The algebraic approach
described in this paper was applied to calculate explicitly the 
effective potential in Yang-Mills theory and to study the
structure of the vacuum of this model
[35]. 

\bigskip
\bigskip
\centerline{\mittel Acknowledgments }
\bigskip

I would like to thank Professor G. Tsagas for his kind 
invitation to present this talk at the International 
Conference `Global Analysis, Differential Geometry and 
Lie Algebras' and H. Christoforidou for the hospitality 
extended to me at the Aristotle University of Thessaloniki.
This work was supported in part by the Alexander von 
Humboldt Foundation.

\bigskip
\bigskip

\centerline{\mittel References}
\bigskip

\item{[1]} B. S. De Witt,
{\it Dynamical Theory of Groups and Fields}, 
(Gordon and Breach, New York, 1965)

\item{[2]} G. A. Vilkovisky, {\it The Gospel according to De Witt},
in: {\it Quantum Theory of Gravity}, 
ed.  S. Christensen (Hilger, Bristol, 1983) p. 169

\item{[3]} A. O. Barvinsky and G. A. Vilkovisky,
Phys. Rep. {\bf C 119} (1985) 1

\item{[4]} I. G. Avramidi, 
{\it Covariant Methods for the Calculation of 
the Effective Action in Quantum Field Theory and Investigation of 
Higher-Derivative Quantum Gravity}, 
PhD thesis, Moscow State University (1986), UDK 530.12:531.51, 178 pp. 
[in Russian]; 
Transl. available at @xxx.lanl.gov\ /hep-th/9510140, 122 pp.

\item{[5]} I. G. Avramidi,
Nucl. Phys. {\bf B 355} (1991) 712

\item{[6]} P. B. Gilkey,
{\it Invariance Theory, the Heat Equation and the  Atiyah - Singer
Index Theorem}, 
(Publish or Perish, Wilmington, 1984)

\item{[7]} J. Hadamard,
{\it Lectures on Cauchy's Problem}, 
in: {\it Linear Partial Differential Equations}, 
(Yale U. P., New Haven, 1923)

\item{[8]} S. Minakshisundaram and A. Pleijel,
Can. J. Math. {\bf 1} (1949) 242

\item{[9]} R. T. Seeley,
Proc. Symp. Pure Math. {\bf 10} (1967) 288

\item{[10]} R. Schimming,
Beitr. Anal. {\bf 15} (1981) 77

\item{[11]} R.  Camporesi,
Phys. Rep. {\bf 196} (1990) 1

\item{[12]} J. S. Dowker,
Ann. Phys. (USA) 62 (1971) 361

\item{[13]} J. S. Dowker,	
J. Phys. A 3 (1970) 451

\item{[14]} S. W. Hawking, 
Comm. Math. Phys. {\bf 55} (1977) 133

\item{[15]} B. S. De Witt, 
{\it Quantum gravity: new synthesis}, 
in: {\it General Relativity}, eds. S. Hawking and W. Israel, 
(Cambridge Univ. Press., Cambridge, 1979)

\item{[16]} I. G. Avramidi, 
{\it Covariant approximation schemes 
for calculation of the heat kernel in quantum field theory}, 
University of Greifswald (September, 1995), 19 pp., Proc. 
Int. Seminar ``Quantum Gravity'', Moscow, June 12--19, to appear;
available at @.xxx.lanl.gov/hep-th/9509075

\item{[17]} I. G. Avramidi,
{\it Covariant methods for calculating the low-energy effective
action in quantum field theory and quantum gravity},
University of Greifswald (March, 1994);
available at @xxx.lanl.gov/ gr-qc/9403036

\item{[18]} I. G. Avramidi,
Phys. Lett. {\bf B 305} (1993) 27

\item{[19]} I. G. Avramidi, 
Phys. Lett. {\bf B 336} (1994) 171

\item{[20]} I. G. Avramidi, 
J. Math. Phys. {\bf 37} (1996) 374

\item{[21]} I. G. Avramidi, 
{\it  New algebraic methods for calculating 
the heat kernel and the effective action in quantum 
gravity and gauge theories},
in: {\it `Heat Kernel Techniques and Quantum Gravity'}, 
ed. S. A. Fulling, 
{\it Discourses in Mathematics and Its Applications},  
(Department of Mathematics, Texas A\& M University, 
College Station, Texas, 1995), pp.~115--140

\item{[22]} I. G. Avramidi, 
J. Math. Phys. {\bf 36} (1995) 5055

\item{[23]} P. B. Gilkey,
J. Diff. Geom. {\bf 10} (1975) 601

\item{[24]} I. G. Avramidi,
Teor. Mat. Fiz. {\bf 79} (1989) 219

\item{[25]} I. G. Avramidi,	
Phys. Lett. {\bf B 238} (1990) 92

\item{[26]} P. Amsterdamski, A. L. Berkin and D. J. O'Connor,
Class. Quantum  Grav. {\bf 6} (1989) 1981

\item{[27]} R. Schimming, 
{\it Calculation of the heat kernel coefficients,}
in: {\it Analysis, Geometry and Groups: A Riemann 
Legacy Volume}, ed. H. M. Srivastava and Th. M. Rassias,
(Hadronic Press, Palm Harbor, 1993), part. II, p. 627

\item{[28]} I. G. Avramidi and R. Schimming, 
{\it Algorithms for the calculation of the heat kernel coefficients},
University of Greifswald (October, 1995), 12 pp., 
Proc. IIIrd Workshop "Quantum Field Theory under the Influence of 
External Conditions", 
Leipzig, 18--22 Sept. 1995, to appear;
available at @xxx.lanl.gov/hep-th/9510206, 

\item{[29]} I. G. Avramidi,
Yad. Fiz. {\bf 49} (1989) 1185

\item{[30]} I. G. Avramidi, 
Phys. Lett. {\bf B 236} (1990) 443

\item{[31]} A. O. Barvinsky, Yu. V. Gusev, G. A. Vilkovisky and 
V. V. Zhytnikov, 
J. Math. Phys. {\bf 35} (1994) 3543

\item{[32]} J. A. Wolf, 
{\it Spaces of Constant Curvature}, 
(University of California, Berkeley, CA, 1972)

\item{[33]} M. Takeuchi, 
{\it Lie Groups II}, 
in: {\it Translations of Mathematical Monographs}, vol. 85, 
(AMS, Providence, 1991), p. 167 

\item{[34]} T. Branson, P. B. Gilkey and B. \O rsted,
Proc. Amer. Math. Soc. {\bf 109} (1990) 437

\item{[35]} I. G. Avramidi, 
J. Math. Phys. {\bf 36} (1995) 1557

\bye